\documentclass[utf8]{FrontiersinHarvard} 

\usepackage{url,hyperref,lineno,microtype,subcaption}
\usepackage[onehalfspacing]{setspace}
\usepackage{multirow}
\usepackage{booktabs}

\def\firstAuthorLast{Villarreal-Haro {et~al.}} 
\def\Authors{Juan Luis Villarreal-Haro$^{1,*}$, Remy Gardier\,$^{1}$, Erick J Canales-Rodríguez$^{1}$, Elda Fischi Gomez$^{2,3,1}$, Gabriel Girard$^{1,2,3,4}$, Jean-Philippe Thiran$^{1,2,3}$, Jonathan Rafael-Patiño$^{3,1}$}
\def\Address{$^{1}$Signal Processing Laboratory (LTS5), École Polytechnique Fédérale de Lausanne (EPFL), Lausanne, Switzerland\\
$^{2}$CIBM Center for Biomedical Imaging, Switzerland\\
$^{3}$Radiology Department, Centre Hospitalier Universitaire Vaudois and University of Lausanne, Lausanne, Switzerland\\
$^{4}$Department of Computer Science, University of Sherbrooke, Sherbrooke, QC, Canada}

\def\corrAuthor{Juan Luis Villarreal-Haro}

\def\corrEmail{juan.villarrealharo@epfl.ch}

\usepackage[norelsize, linesnumbered, ruled, lined, boxed, commentsnumbered, noend]{algorithm2e}

\usepackage{float}

\usepackage{amsthm}
\theoremstyle{definition}
\usepackage[framemethod=TikZ]{mdframed}
\usepackage{lipsum}
\mdfsetup{%
middlelinewidth=1pt,
roundcorner=10pt}
\newmdtheoremenv{definition}{Definition}[section]
\newcommand\norm[1]{\lVert\mathbf{#1}\rVert}
\newcommand\simplenorm[1]{\lVert #1\rVert}

\DeclareMathOperator{\argmin}{arg\,min} 
\usepackage{xcolor}

\SetCommentSty{mycommfont}

\begin{document}
\onecolumn
\firstpage{1}

\title[Computational Axonal Configurator]{Computational Axonal Configurator for Tailored and Utradense Substrates (CACTUS)}

\title[Generating Realistic White Matter Substrates with CACTUS]{CACTUS: A Computational Framework for Generating Realistic White Matter Microstructure Substrates} 

\author[\firstAuthorLast ]{\Authors} 
\address{} 
\correspondance{} 

\extraAuth{}

\maketitle

\begin{abstract}

Monte-Carlo diffusion simulations are a powerful tool for validating tissue microstructure models by generating synthetic diffusion-weighted magnetic resonance images (DW-MRI) in controlled environments. This is fundamental for understanding the link between micrometre-scale tissue properties and DW-MRI signals measured at the millimetre-scale, optimising acquisition protocols to target microstructure properties of interest, and exploring the robustness and accuracy of estimation methods. However, accurate simulations require substrates that reflect the main microstructural features of the studied tissue. To address this challenge, we introduce a novel computational workflow, CACTUS (Computational Axonal Configurator for Tailored and Ultradense Substrates), for generating synthetic white matter substrates. Our approach allows constructing substrates with higher packing density than existing methods, up to 95 $\%$ intra-axonal volume fraction, and larger voxel sizes of up to $(500 \mu m)^3$ with rich fibre complexity. CACTUS generates bundles with angular dispersion, bundle crossings, and variations along the fibres of their inner and outer radii and g-ratio. We achieve this by introducing a novel global cost function and a fibre radial growth approach that allows substrates to match predefined targeted characteristics and mirror those reported in histological studies. CACTUS improves the development of complex synthetic substrates, paving the way for future applications in microstructure imaging.

\tiny
\textbf{keywords}: microstructure imaging, diffusion MRI, brain imaging, white matter, Monte Carlo simulations, numerical phantom, synthetic substrates, high packing density.
\end{abstract}

\section{Introduction}

Diffusion-weighted magnetic resonance imaging (DW-MRI) is a non-invasive technique used to study the microscopic structure of biological tissues in vivo. It is sensitive to the ensemble of water molecules (wherein each molecule follows a random motion pattern) as they interact with cellular surfaces~~\citep{bihan1995molecular}. ~\citep{bihan1995molecular}. This technique provides a valuable tool to study brain microstructure and its alterations following injury~\citep{parizel2005new,to2022subacute} and neurological disease~\citep{VanGelderen1994,budde2010neurite,narvaez2019histological}.

White matter is a crucial component of the brain, composed of highly organised axon bundles that interconnect cortical regions and subcortical regions~\citep{bruckner1996extracellular, sporns2011non}. Various imaging techniques have been considered to characterise the white matter tissue microstructure in different species. For example, axon diameters have been measured in some white matter regions of the macaque monkey brain using histology and DW-MRI ~\citep{caminiti2013diameter}, and optical microscopy~\citep{innocenti2017axon}. These studies show that the estimated distribution of axon diameters is long-tailed, with a mean of around one micrometre. A recent study using high-resolution three-dimensional (3D) synchrotron X-ray nano-holotomography found that axons are non-cylindrical and exhibit environment-dependent variations in diameter and trajectory~\citep{andersson2020axon}. Alongside axon diameters, another relevant feature is the intracellular volume the axons occupy in a predetermined region. In histological postmortem data, the white matter intracellular space volume has been estimated as ranging between 60-85\% of the brain volume for macaques~\citep{stikov2015vivo} and human adults~\citep{sykova2008diffusion}. Interestingly, it goes as high as 70\%-95\% in mice, as reported by light microscopy~\citep{tonnesen2018super}, and cryo and chemical fixations~\citep{korogod2015ultrastructural}.

Given the importance of studying white matter tissue microstructure in vivo, several DW-MRI models have been proposed~\citep[e.g.][]{Jelescu2017, Novikov2018, Novikov2019, Assaf2004, Assaf2008, ALEXANDER2010, Dyrby2011, Veraart2021b, Neuman1974, VanGelderen1994, Murday1968, Soderman1995, LEE2020}. However, validating these non-invasive techniques requires physical and numerical phantoms with a well-known microstructure~\citep{Lavdas2013, Campbell2005, Fillard2011, Tournier2008, Fieremans2008, Schilling2019c, zhou2018axon, maier2017challenge, andersson2020axon, LEE2020, rafael2020robust}. Phantoms, in the context of this paper, are geometrical models of brain tissue structures that serve as a proxy or reference for evaluating the performance of imaging techniques. While physical phantoms have been widely used, they are often limited by their high costs and the impracticality of replicating axons' sizes and complex spatial arrangement. Therefore, numerical phantoms have emerged as the most popular validation technique for studying the complexities of diffusion phenomena in cases where analytical solutions are unavailable; because they only require a substrate that mimics the tissue of interest to simulate the displacements of water molecules and corresponding DW-MRI signal~\citep{Close2009, Neher2014, Cote2013}. Nevertheless, the difficulty in Monte-Carlo simulations lies in accurately mimicking the geometry of white matter tissue~\citep{kerkela2021comparative,rafael2020robust,hall2009convergence,nilsson2017resolution,baxter2013computational,plante2013monte,nilsson2012importance, truffet2020evolutionary}.

Various studies have attempted to generate numerical phantoms approaching the tissue's morphological complexity and density. For instance, two popular tools, MEDUSA~\citep{Ginsburger2019} and CONFIG~\citep{Callaghan2020}, focus on generating specialised voxel-wise phantoms with microstructural geometries that replicate the properties of white matter. Recently, a tailored modification of~\cite{Close2009} framework was used to build challenging substrates for the DiSCo challenge~\citep{rafael2021microstructural}, aimed to test fibre-tracking and connectivity methods on large-scale synthetic datasets from DW-MRI Monte-Carlo simulations. While these methods have provided valuable tools to characterise and simulate DW-MRI signals in numerical substrates, they still have important limitations regarding the maximum packing density and substrate size achieved. For instance, state-of-the-art frameworks can generate synthetic substrates with packing densities up to 75\%~\citep{Callaghan2020, Ginsburger2019, rafael2021microstructural}, whereas the density found in histological data goes up to 95\% in some regions~\citep{korogod2015ultrastructural,tonnesen2018super}. Moreover, they cannot sample substrate beyond $(100 \mu m)^3$, which in turn restricts the sampling diversity achieved for morphological features~\citep{rafael2020robust,romascano2018voxel}. Therefore, the DW-MRI signals generated from these substrates may not accurately mimic the brain signals measured in white matter regions with higher packing densities.


To overcome these limitations, we introduce a novel computational workflow, CACTUS (Computational Axonal Configurator for Tailored and Ultradense Substrates), to generate synthetic fibres with rich microstructure characteristics. Expanding on previous methods~\citep{Close2009, Ginsburger2019, rafael2020robust}, we develop a novel numerical phantom generator for white matter substrates. CACTUS solves the high-density packing problem and achieves up to 95\% intracellular volume fractions while efficiently generating substrate sizes up to $(500 \mu m)^3$. Furthermore, CACTUS is highly customisable, capable of generating synthetic substrates with a wide range of characteristics, such as single-bundle~\citep{stikov2015vivo}, bundle crossings~\citep{schilling2017can, TOURNIER2007, Tuch2004, CanalesRodriguez2019}, orientation dispersion~\citep{zhang2012noddi, Daducci2015}, gamma-distributed axon radii~\citep{sepehrband2016parametric, Assaf2008}, non-constant longitudinal fibre-radii~\citep{andersson2020axon}, substrates with non-cylindrical fibres and tortuous surfaces~\citep{lee2019along}, and myelin compartments~\citep{stikov2015vivo, Mackay1994, CanalesRodriguez2021}. Through these features, CACTUS expands on the capabilities of existing substrate generation methods, providing a flexible and versatile tool for studying white matter microstructure in controlled environments.

\section{Methods}

\begin{figure}
\begin{center}
\includegraphics[width=.95\textwidth]{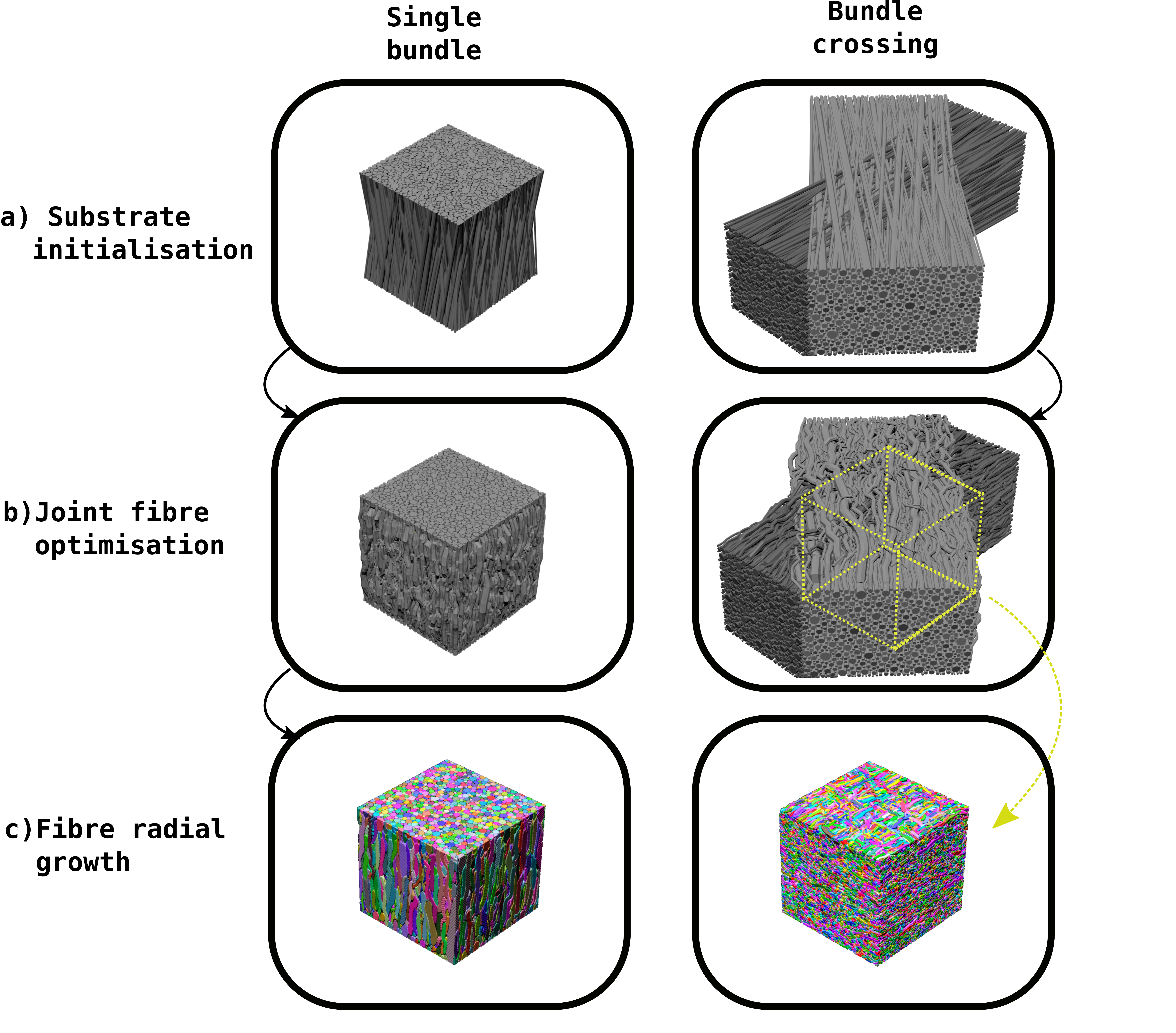}
\end{center}
\caption{Example of the CACTUS method steps to create a synthetic substrate. a) The \textbf{substrate initialisation} orients fibres in a bundle to achieve a predefined mean angular dispersion (e.g., 15º). b) The \textbf{joint fibre optimisation} step removes fibre overlaps by adapting their trajectories and local radii. In the case of bundle crossings, the trajectories are trimmed to the centre of the crossing. c) The \textbf{fibre radial growth} step further increases the fibre-packing density while keeping the predefined target radius distribution.
 }
\label{fig:pipeline}
\end{figure}

CACTUS generates synthetic substrates in three steps (see Figure \ref{fig:pipeline}): \textbf{a) substrate initialisation, b) joint fibre optimisation, c) fibre radial growth (FRG).}

Firstly, in the \textbf{substrate initialisation} step, synthetic straight cylindrical fibres are initialised and parameterised inside a cuboid. In CACTUS, a single fibre population (bundle) is a group of fibres arranged cohesively along one main orientation. A bundle has two main properties: the average global dispersion, which is the mean angle between the main orientation of each fibre and the bundle, and the target radii distribution, from which the fibre radii are sampled. 

In the second step, the \textbf{joint fibre optimisation}, CACTUS extends previously proposed frameworks~\citep{Close2009, Ginsburger2019} based on local optimisation. In our case, we aim to minimise a cost function that penalises some essential fibre properties such as overlapping, high curvature, increase in length, and promote compactness. Moreover, CACTUS introduces a new fibre parameterisation based on capsules, which reduces the number of parameters needed to characterise fibre trajectories and handles fibre overlapping more efficiently. The resulting optimisation problem is solved via a gradient descent algorithm~\citep{duchi2011adaptive}. During optimisation, CACTUS prioritises removing fibre overlapping, while the penalisation of curvature, length and promotion of compactness maintains a coherent fibre structure at all time-points.

Finally, the fibre trajectories are used to mesh the fibre surfaces in the \textbf{fibre radial growth} (FRG) step. The FRG also increases the packing density while keeping the correspondent fibre's parameterisation structure using a discrete grid to seed, to grow, and to rearrange the fibre into the final substrates. The grid discretisation defines the fibres' isosurface needed to compute the final surfaces with a marching cube algorithm~\citep{lewiner2003efficient-marchingCubes}.


\subsection{Substrate initialisation}

\label{subsec:initialisation}

Our substrate initialisation algorithm enhances the circle two-dimensional (2D) packing algorithm proposed by~\citet{Hall2009ConvergenceMRI} to create a 3D packing of bundles. The algorithm creates a single bundle by initialising the fibres inside a cuboid of dimensions $L\times L \times H$. The endpoints of the fibres are contained within the $L\times L$ squared faces, while the orientation of the cuboid's height $H$ and the bundle are aligned to the Z-axis. The algorithm packs 2D circles in the opposite faces of the cuboid, using radii from a gamma $\Gamma(\alpha,\beta)$ distribution~\citep{sepehrband2016parametric, Assaf2008}, until the target density is met. At the same time, the algorithm packs the two opposite 2D circles to create an initialisation such that the bundle reach the specified mean angular dispersion $\boldsymbol{\eta}$.
In order to create a substrate with two bundles crossing at an inter-bundle angle of $\boldsymbol{\theta}$, two different bundles are initialised in their respective cuboids and subsequently rotated and translated are applied. Figure \ref{fig:pipeline} shows examples of a single bundle and a bundle crossing initialisation. 
Finally, we parameterise each fibre's skeleton as the trajectory of its centre of mass. This trajectory is defined by several control points connecting the two endpoints sampled during the packing algorithm, where each point has a corresponding radius.

\subsection{Joint fibre optimisation}
 Once the substrate is initiated as described in subsection \ref{subsec:initialisation}, fibres may overlap. CACTUS employs an optimisation method to readjust the fibre trajectories and disentangle overlaps by defining several cost functions. 
 These cost functions, inspired by~\citet{Close2009} and~\citet{Ginsburger2019}, help to regularise and obtain coherent fibre structures with the specified target properties. Ordered by priority of penalisation, these target properties are as follows: i) fibre overlapping (see subsection \ref{subsub:overlap_cost}), ii) high curvature, iii) increased fibre length,  iv) changes in radii and v) compactness. 
 
In the following subsection, we introduce the novel parameterisation and overlapping cost function based on capsules, which is a key contribution of our work. As the remaining cost functions are relatively straightforward and similar to those in previous studies, we have provided their definitions in Supplementary Material (section joint fibre optimisation).

\subsubsection{Fibre capsule-parameterisation}
\begin{figure}
    \centering
	\includegraphics[width = .8\textwidth]{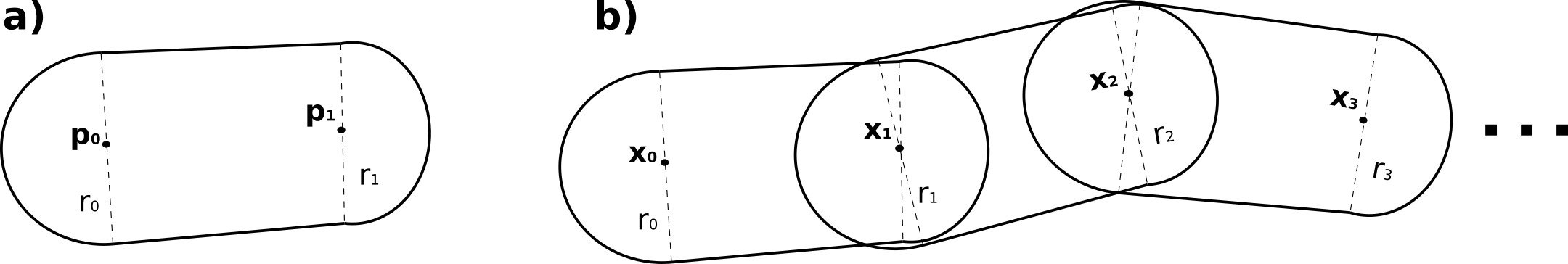}  \hfill
    \caption{a) Capsule example, whose parameters are two points and a radius at each control point. b) Example of fibre as a chain of capsules. Two adjacent capsules share a control point (position and radius parameters).}
    \label{fig:capsule_definition}
\end{figure}

Fibres are parameterised as skeletons made of 3D control points. In the overlapping cost function, every pair of consecutive points in the skeleton forms a \textbf{capsule}, defined with the set of parameters $\left[ \mathbf{p_0}, \mathbf{p_1}, r_0, r_1 \right]$, where $\mathbf{p_0}, \mathbf{p_1} \in \mathbb{R}^3$ are the initial/ending points of the capsule, and $r_0, r_1 \in \mathbb{R}$ are their respective radius (see Figure \ref{fig:capsule_definition} a).
In this framework, a fibre parameterisation can be defined as a \textbf{chain of capsules} (see Figure \ref{fig:capsule_definition} b). The fibre $\mathcal{S}^a$, with $m_a$ control points, is composed of the capsules determined by the subsequent point pairs as $\left[ \mathbf{x_i^a}, \mathbf{x_{i+1}^a}, r_i^a, r_{i+1}^a \right]$ with control points $\left\{ \mathbf{x_0}^a, \mathbf{x_1}^a, \dots \mathbf{x_{m_a-1}}^a \right\} \subset \mathbb{R}^3$ and associated radius $ \left\{  r_0^a, r_1^a, \dots r^a_{m_a-1} \right\} \subset \mathbb{R} $.   

\subsubsection{Overlapping cost function}
\label{subsub:overlap_cost}

The \textbf{overlapping cost function} handles the fibre collision by identifying overlaps from two capsules from two different fibres. In CACTUS, the detection step of capsule intersection is a generalisation of the cylinder-to-cylinder collision detection~\citep{van2015essential}. We define the overlapping cost function between two capsules by computing the overlapping of the closest spheres centred in the capsules. Formally, the closest points between two given capsules $[\mathbf{p_0,p_1},r_{p_0},r_{p_1}]$, $[\mathbf{q_0,q_1}, r_{q_0}, r_{q_1}]$, with $\mathbf{p_i , q_j} \in \mathbb{R}^3$ and $, r_{p_i}, r_{q_j} \in \mathbf{R}$, are the points centred in the capsule $\left( (1-t_p)\mathbf{p_0} + t_p \mathbf{p_1} \right)$ and, $( (1-t_q)\mathbf{q_0} + t_q \mathbf{q_1} )$, where $t_p, t_q$ are found by the following minimisation problem:
	\begin{align}
		g(t_a , t_b ; \mathbf{p_0 , p_1 , q_0 , q_1} ) & =
		\simplenorm{(1-t_a)\mathbf{p_0} +
			t_a\mathbf{p_1} - (1-t_b)\mathbf{q_0} - t_b\mathbf{q_1} }^2
		\\
		t_p , t_q := \operatorname*{\argmin}\limits_{\substack{t_a,t_b
			\\ s.t. \; 0 \leq t_a,t_b \leq 1}} &  g(t_a , t_b ;
		\mathbf{p_0 , p_1 , q_0 , q_1 }),
		\label{eq:min_distances}
	\end{align}
	which has a closed-form solution.

After finding the values $t_p, t_q$ that define the closest centre points between two capsules of different fibres, their overlapping cost function is defined as:
\begin{equation}
 f_{1} \left(\mathbf{p_0 , p_1 , q_0 , q_1} ,r_{p_0}  , r_{p_1}, r_{q_0} , r_{q_1}  ; t_p,t_q \right) = 
    \begin{cases}
        D^2 \norm{p_0 - p_1} \norm{q_0 -q_1}  r_p r_q , & \text{if } D \geq 0\\
       0 , & \text{if } D < 0
    \end{cases}
\end{equation}
 where, 
 \begin{align}
 D &:= \left(1 -\frac{\norm{c_1 - c_2}} {r_p + r_q} \right), \\
 c_1 & := (1-t_p)\mathbf{p_0 } +t_p \mathbf{p_1} , \\
 c_2 & :=		(1-t_q)\mathbf{q_0} + t_q\mathbf{q_1} ,\\
 r_p &:= (1-t_p)r_{p_0} + t_p r_{p_1},\\
 r_q &:= (1- t_q)r_{q_0} + t_q r_{q_1}, \\
 \end{align}
 
and $t_p, t_r$ are the minimal values from the function in Eq.~\ref{eq:min_distances}.

Consequently, the total overlapping cost function in a substrate is computed by adding the evaluated cost of all possible pairwise capsule combinations. If capsules overlap, a penalisation is added; otherwise, it is set to zero.

\subsubsection{Implementation details}

At last, we mention the technical implementation details of the joint-fibre optimization algorithm, including strategies for reducing computational complexity and the use of specific data structures. Firstly, in the total overlapping cost function, the capsule-to-capsule comparison is a $\mathcal{O}(n^2)$ problem. To improve computational time, we implemented a fixed-radius-cell data-structure~\citep{turau1991fixed} for nearest neighbours queries, reducing the problem to $\mathcal{O}(n)$. Since all the cost functions are analytical, we calculated their analytical derivatives for the gradient descent algorithm (see Supplementary materials, section joint fibre optimisation). 
We used the adaptative gradient Adagrad~\citep{duchi2011adaptive}, iterating until there were no overlapping fibres.
All the cost functions, queries, and gradients calculations were implemented in C++~\citep{stroustrup1999overview} and parallelised with OpenMP~\citep{chandra2001parallel}. 
To handle bundle crossings, we trim the optimised fibre trajectories to keep only a subregion with fibres that truly belong to the crossing, as shown in Figure \ref{fig:pipeline}. This step eliminates boundary fibres that may not fully represent the crossing characteristics.

\subsection{Fibre radial growth (FRG)}
\subsubsection{FRG description}
After completing the substrate initialisation and joint fibre optimisation steps, it follows to compute the fibre mesh. Previous studies have managed to achieve a fibre density up to 75\% ~\citep{Ginsburger2019,mingasson2017axonpacking,altendorf2011random}  with cylindrical-shaped fibres and gamma-distributed diameter. In this study, we propose a new method, called Fibre Radial Growth (FRG), to obtain higher packing density and complex axon morphologies beyond the cylindrical shape. The FRG algorithm discretises the 3D space that the fibres occupy to define individual masks for each fibre in it. The FRG algorithm begins to generate the fibre masks by randomly placing seed points within all capsule fibres. These seed points grow iteratively by adding neighbouring points to the fibre mask, employing a breadth-first-search approach through the grid. The seeds grow for a fixed number of iterations as long as they do not interfere with other fibres' boundaries. The propagation through random initialisations avoids uniform growth and adds irregularities to the fibre shape, allowing tortuous surface reconstructions in the fibre surfaces. Since the seeding is done inside capsules, the final axon radius in the mesh is related to the radii used in the capsules.

Once the FRG step is completed, the fibre density of the particular configuration inputted is maximised. We compute the fibre's outer surface mesh using the fibre masks with the Lewiner marching cubes algorithm~\citep{lewiner2003efficient-marchingCubes,scikit-learn}. Then, we applied a Laplacian smoothing~\citep{sorkine2004laplacian,herrmann1976laplacian,sullivan2019pyvista} to remove sharp angles, and finally decimate the mesh to reduce the number of triangles without affecting the morphology of the substrates~\citep{shekhar1996octree}. Subsequently, we generate a new mesh representing the fibre's inner surface by eroding the previously estimated outer grid and following the same procedure for the meshing. The space between these two surfaces defines the myelin volume.

\subsubsection{Implementation details}

Finally, we would like to elaborate on the technical implementation details of the FRG algorithm to mention the specific design choices we made to ensure its computational efficiency. FRG is implemented in Python~\citep{python}, parallelised with its multiprocessing ibraries~\citep{python_multiprocessing}, and compiled with Numba~\citep{numba_compile}. Image 3D processing and meshing are done using~\citet{scikit-image, sullivan2019pyvista,Hess_blender}. 
Moreover, the FRG is designed to run a ball-tree structure~\citep{moore2003new_balltree} from the Sklearn library~\citep{scikit-learn} as a preprocessing to store fibres and their interactions. The fine-tuned FRG algorithm's design allows for the independent execution of fibre growth and meshing on multiple computers in a distributed manner, eliminating the need for multi-thread or computer synchronisation.

\section{Experiments}

To evaluate the performance of CACTUS, we designed a comprehensive set of substrates with specific geometries. Each experiment below involves several metrics essential for quantifying the microstructure properties of the brain white matter. The metrics include the axon volume fractions, the radius distribution per substrate, the radii change along the fibres, the myelin volume, the g-ratio, the orientation dispersion and bundle crossings.

\subsection{Maximum fibre volume fraction}

In our first experiment, we aim to explore the macro-structural parameters of substrates, such as substrate size (i.e. the voxel size in MRI experiments), fibre dispersion, two bundle crossings, and the ability to create high-density packing substrates. We assess the maximum fibre volume fraction that CACTUS achieves in two scenarios: a single bundle and two bundles. In the single bundle case, we generated six substrates with mean angle dispersions of $0^\circ$, $5^\circ$, $10^\circ$, $15^\circ$, $20^\circ$, and $25^\circ$, respectively. In the two bundles case, we generated five crossing substrates with inter-bundle angles of $30^\circ$, $45^\circ$, $60^\circ$, $75^\circ$, and $90^\circ$, and the fibres of each bundle were initialised with a mean angle dispersion of $5^\circ$ around the main bundle orientation.

 \subsection{Substrates targeting predefined microstructure features}
 The following two paragraphs describe experiments conducted to explore the ability of CACTUS to replicate desired microstructural parameters into its synthetic substrates. These parameters include the axon volume fraction (AVF), myelin volume fraction (MFV), g-ratio, and radii distribution. We compare the reference values taken from previous histological studies and those achieved by CACTUS.


In the second experiment, we created a series of synthetic substrates that emulate the histological values reported by~\citet{stikov2015vivo} in various white matter regions. Specifically, the target characteristics are the fibre volume fraction, myelin volume fraction, and aggregated g-ratio, $g = \sqrt{1 - MVF/FVF}$~\citep{stikov2015vivo}. In our scenario, the axon volume fraction (AVF) is the volume of the fibre inner surface. The myelin volume fraction (MVF) represents the volume of the space between the inner and outer fibre surfaces. The fibre volume fraction (FVF) is the sum of AVF and MVF.



In the last experiment, we investigated the effect of substrate size on radii distribution. To measure the radii distribution, for each fibre, we cut the mesh skeleton in an orthogonal plane at regular $1 \mu m$ intervals and calculated the cross-sectional area of the polygon defined by the plane. The equivalent fibre radius is defined as the radius of a circle with the same area as the polygon~\citep{lee2019along}. The global radii distribution per substrate was computed using the mean radius for each fibre. 

\section{Results}


\subsection{Maximum fibre volume fraction}

\begin{figure}[t]
    \centering

    \includegraphics[width=.9\textwidth]{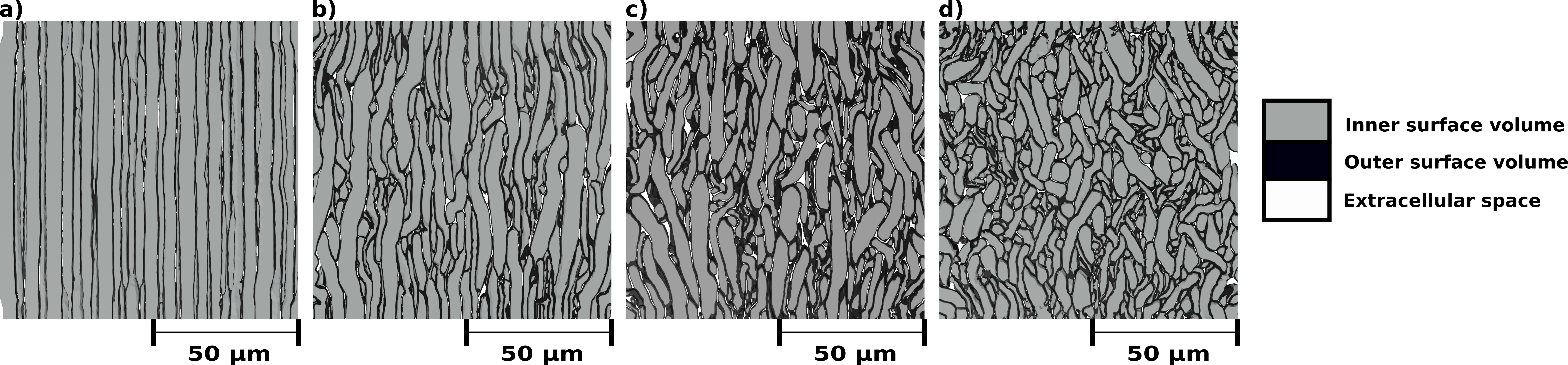}
    \caption{(a-d) Mesh's renders of cross-sections with dimensions of $(100 \mu m)^2$ to visualize the internal morphology of four substrates with a single bundle, each with different mean angular dispersion. For all cases, the outer surface volume is coloured black. The inner surface volume is superimposed over the outer volume and coloured grey. White represents the extracellular space, ie. the volume not occupied by any fibre.
    All bundles are vertically aligned, and the substrates were built to have a mean angular dispersion of a) $0^\circ$, b) $5^\circ$, c) $10^\circ$, and d) $20^\circ$, respectively.}
    \label{fig:substrate_dispersion}
\end{figure}

Figure \ref{fig:substrate_dispersion} shows the internal morphology of four substrates consisting of a single bundle with a dispersion of $0^\circ$, $5^\circ$, $10^\circ$, and $20^\circ$, respectively. All the substrates were generated with dimensions of $(500 \mu m)^3$. Table \ref{tab:dispersion_crossings} (top panel) reports the substrate characteristics, including the number of fibres, the obtained fibre volume fraction, and the dispersion parameters. We note that the maximum fibre volume fraction decreased from $~94.7\%$ to $90.8\%$ as the dispersion increased from $0^\circ$ to $25^\circ$.

\begin{figure}[H]
    \centering
    \includegraphics[width= .9\textwidth]{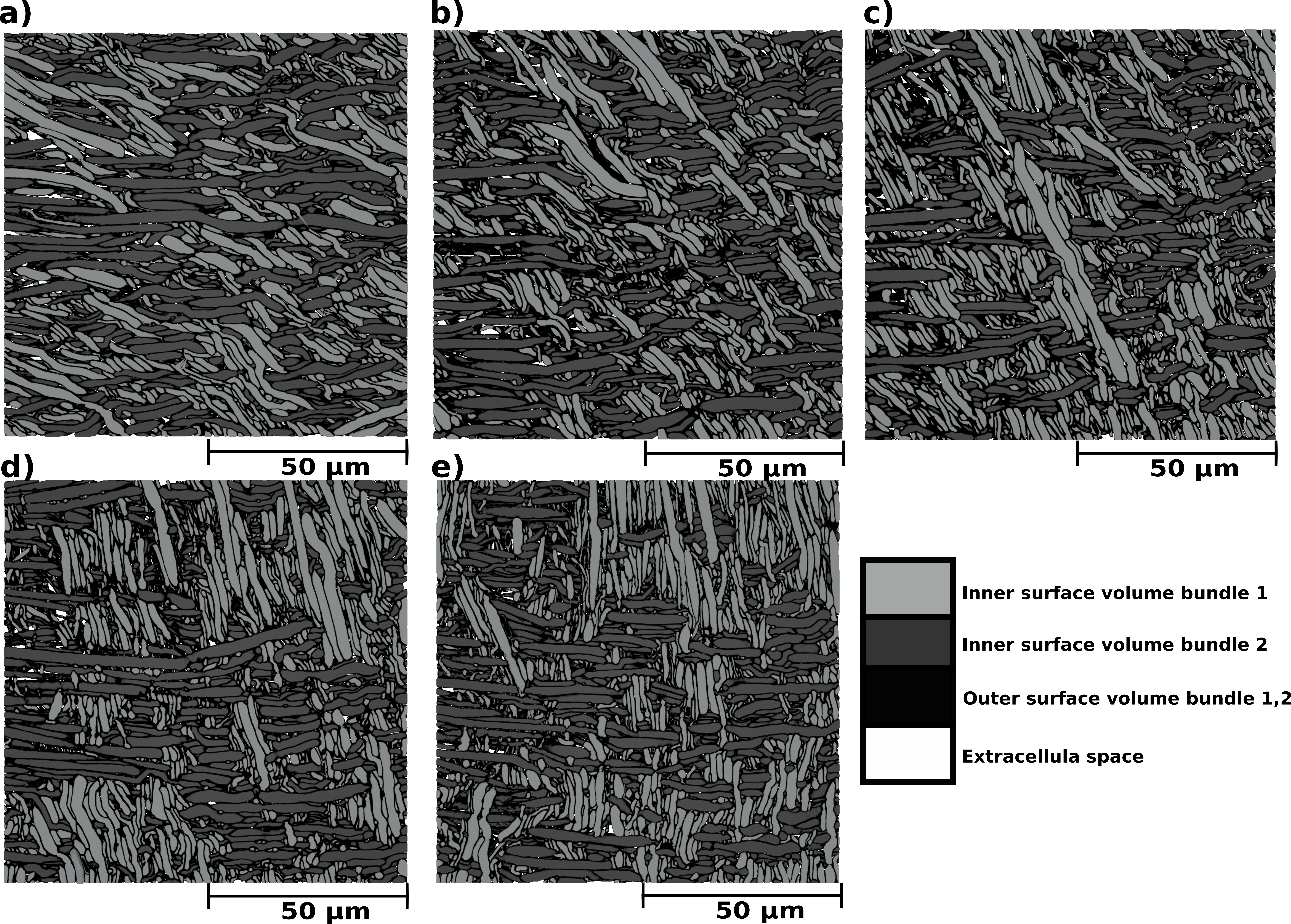}
    \caption{
    (a-e) Mesh renders of cross-sections with dimensions of $(100 \mu m)^2$ portraying the internal morphology of five substrates consisting of two bundles with different inter-bundle angles. In all cases, the outer surface volume of bundle 1 and bundle 2 is displayed in black. The inner surface volume is superimposed over the outer volume and coloured light grey for bundle 1 and dark grey for bundle 2. The extracellular space, representing the volume not occupied by any fibre, is coloured white.
 The dark grey fibres are aligned parallel to the $X$ axis, and the light grey fibres crossed at angles of a) $30^\circ$, b) $45^\circ$, c) $60^\circ$, d) $75^\circ$, and e) $90^\circ$.  The outer volume, which is defined by the outer surfaces minus the inner volume, is coloured in black for both bundles. Extra axonal space is coloured in white.}
    \label{fig:substrate_crossings}
\end{figure}

Results from the experiment generating bundle crossings with different inter-bundle angles are depicted in Figure \ref{fig:substrate_crossings} and Table \ref{tab:dispersion_crossings} (bottom panel). Figure \ref{fig:substrate_crossings} displays a cross-section of the substrates, where each bundle has a distinctive colour for visualisation purposes. Although local perturbations in fibre trajectories (on the order of $5^\circ$) may occur in the substrates due to the high fibre packing, the average bundle orientation is sustained. The bottom panel of Table \ref{tab:dispersion_crossings} reports the fibre volume fraction of these bundle crossing substrates with inter-bundle angles of $30^\circ$, $45^\circ$, $60^\circ$, $75^\circ$, and $90^\circ$. For all the evaluated substrates, the fibre volume fraction remains nearly constant at approximately $93\%$ ($92.2 \%-93.9\%$).

\begin{table}[H]
    \centering
\begin{tabular}{c c c cc} 
 Nbr of Bundles & Bundle dispersion $(\eta)$ & Crossing angle $(\theta)$ & Nbr of fibres  & FVF \\
 \hline
 1 & $0^\circ$   & - & 31,954 & 94.7\%\\
 1 & $5^\circ$   & - &  31,023 & 93.4\%\\
 1 & $10^\circ$  & - & 30,241  & 92.6\%\\
 1 & $15^\circ$  & - & 30,412  & 92.2\%\\
 1 & $20^\circ$  & - & 31,161 & 91.8\%\\
 1 & $25^\circ$  & - & 31,863  & 90.8\%\\
 \hline
2 & $5^\circ$& $30^\circ$  & 30,026 & 93.9\%\\
2 & $5^\circ$& $45^\circ$  & 30,712  & 93.3\%\\
2 & $5^\circ$& $60^\circ$  & 31,023 & 93.5\%\\
2 & $5^\circ$& $75^\circ$  & 31,152 & 92.3\%\\
2 & $5^\circ$& $90^\circ$  & 30,245,  & 92.2\%\\
\end{tabular}
\caption{Substrate characteristics, including the number of bundles, mean dispersion angle, mean inter-bundle crossing angle, number of fibres per substrate, and fibre volume fraction (in per cent), respectively. The top and bottom panels correspond to the substrates with a single bundle and two bundles. Each row represents a different substrate.}
\label{tab:dispersion_crossings}
\end{table}

\subsection{Substrates targeting predefined microstructure features}

\subsubsection{Axon volume fraction, myelin volume fraction, and g-ratio}

\begin{figure}
    \centering
    \includegraphics[ width =\textwidth]{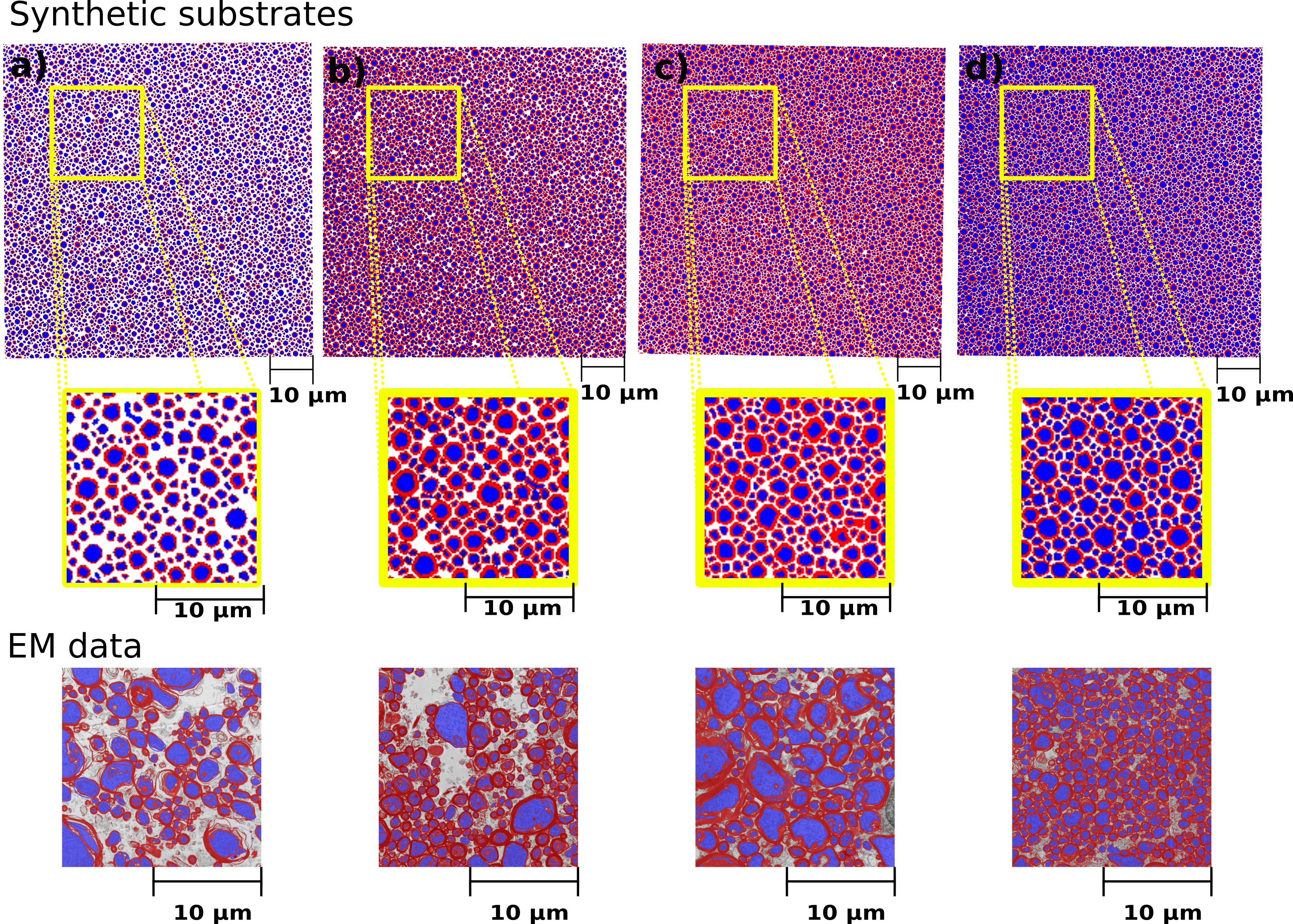}
    \caption{Cross-sections of the synthetic substrates constructed to match the statistics of the histological values reported in Table \ref{tab:target_AVF}. The substrate dimension is $(300 \mu m)^3$. For visualisation purposes, the axonal space is coloured blue, and the myelin is red, and extra axonal space is coloured white. Panels a) - d) correspond to the same substrates shown in Table \ref{tab:target_AVF}. The bottom panel shows the representative histological images courtesy of Prof. Nikola Stikov and Dr. Jennifer Campbell. The EM images are used to show the similarities in fibre shape and packing.}
    \label{fig:stikov_substrate}
\end{figure}

We simulated various substrates of a single bundle to mimic microstructure properties previously reported in~\citet{stikov2015vivo}. The histological values used as a reference are the myelin volume fraction (MVF), fibre volume fraction (FVF), axonal volume fraction (AVF=FVF-MVF), and g-ratio. 
The values achieved by CACTUS are shown in Table \ref{tab:target_AVF}. The difference between the target and obtained substrate properties was lower than $2\%$ in all cases. Examples of the generated substrates and histology data are shown in Figure \ref{fig:stikov_substrate}. Electron microscopy images were generously provided by Prof. Nikola Stikov and Dr. Jennifer Campbell, and are used to highlight the geometric similarities of synthetic fibre shapes.

\begin{table}[H]
\centering
\begin{tabular}{ c  c c c c   c c c c  } \toprule

 & \multicolumn{4}{c }{Target} & \multicolumn{4}{c }{Achieved}\\ 
  \cmidrule(lr){2-5} \cmidrule(lr){6-9}
Substrate   &AVF & MVF &   FVF &  g-ratio &  AVF &  MVF  &  FVF  & g-ratio \\
 \cmidrule(lr){1-1}  \cmidrule(lr){2-5} \cmidrule(lr){6-9}
a) & 25  & 	35  & 	60 & 64.5  & 26.0  & 36.0 &  62 &   64.7   \\ 
b) & 25  & 	43  & 	68 & 60.6   & 26.3 & 43.6 & 69.9    &  61.3  \\    
c) & 31  & 	44  & 	75 &  64.2  & 32.2 & 43.8  &  76.07 & 65    \\   
d) & 39  & 	37  & 	76 &  71.6   & 41.2 & 35.0  &  76.0  &  73.5  \\     
\bottomrule
\end{tabular}
\caption{Target microstructure histological properties (left) reported in~\citet{stikov2015vivo}, and corresponding properties of the substrates generated by CACTUS (right). The axon volume fraction (AVF) is the volume of the inner axon surface. The myelin volume fraction (MVF) represents the volume of the space between the inner and outer axon surfaces. The fibre volume fraction (FVF) is the sum of AVF and MVF. The aggregated g-ratio, $g = \sqrt{1 - MVF/FVF}$ is equal to the mean inner and outer axon radius ratio for all the fibres in the substrate.}
\label{tab:target_AVF}
\end{table}

\subsubsection{Radii distribution and substrate size}

\begin{figure}
    \centering
    \includegraphics[width = \textwidth]{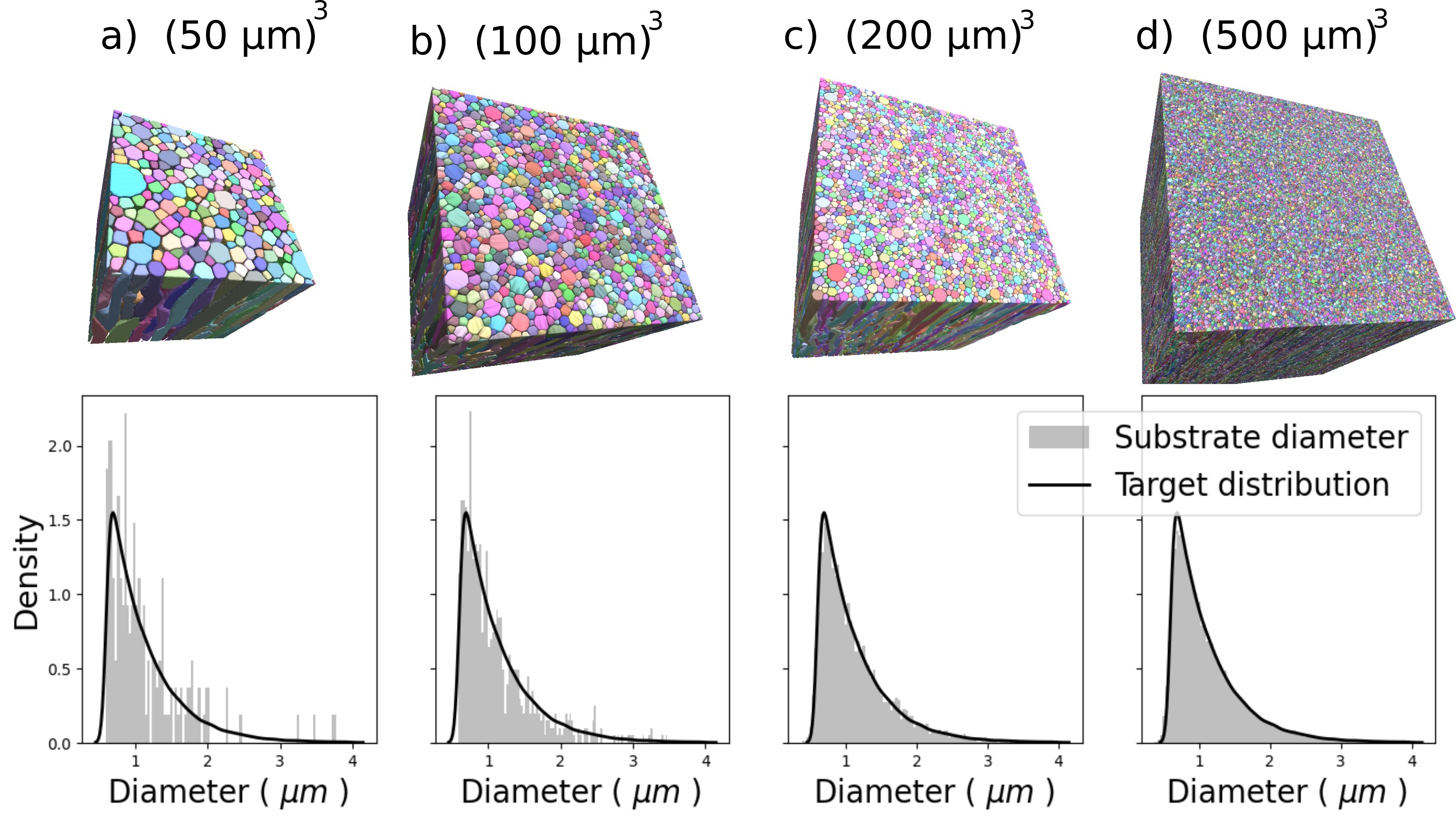}
    \caption{Four 3D substrates of varying sizes: a) $50^3$, b) $100^3$, c) $200^3$, and d) $500^3$ $(\mu m)^3$, with 341, 1316, 4859, and 33478 fibres, respectively. The empirical and target radii distributions are displayed on the bottom of each substrate. The empirical distribution better approximates the target distribution as substrate size increases.}
    \label{fig:substrate_sizes}
\end{figure}

Figure \ref{fig:substrate_sizes} show the
CACTUS substrates with different sizes, ranging from $(30 \mu m)^3$ to $(500 \mu m)^3$, and the target and empirical radius distributions obtained for each substrate. The empirical radius distributions closely replicated the targeted ones for substrates equal to or bigger than $(200 \mu m )^3$. The optimisation algorithm step ran for approximately 4 hours for the largest substrate (right panel) on a node with 64 cores (2.4 GHz) and 400 Mb of RAM. The reconstruction time of the FRG algorithm was approximately one minute per fibre, using one core with 500 Mbs of memory per core.

\begin{figure}
    \centering
    \includegraphics[width = .45 \textwidth]{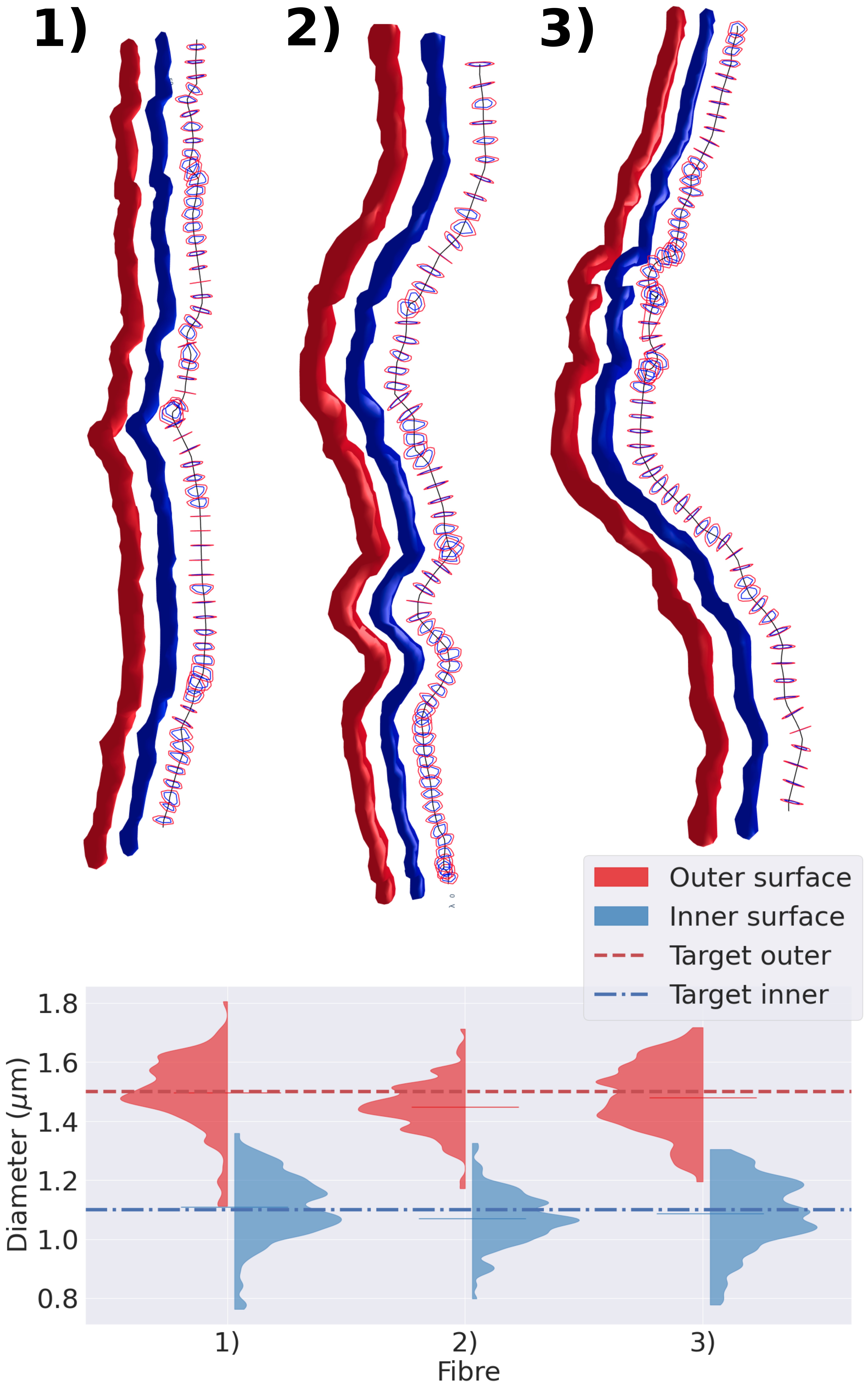}

    \caption{Representative group of three fibres extracted from the substrate shown in Figure \ref{fig:stikov_substrate}. In the top panel, we display the fibres with varying diameters and tortuous trajectories. The straightest fibre is presented on the left, while the most tortuous one is displayed on the right. We display each fibre's outer surface in red and its inner surface in blue. We also show the fibre's skeleton in black and cross-sections orthogonal to the fibre's skeleton. The diameter is measured every $1 \mu m $ along its trajectory. The cross-section cut of the outer surface is shown in red, and the cross-section cut of the inner surface is shown in blue. The bottom panel presents the violin plots of the outer and inner diameters measured. The red (blue) dotted line represents the target outer (inner) diameter of the three fibres.}
    \label{fig:single_fibres}
\end{figure}

We extracted three representative fibre segments from the substrates shown in Figure \ref{fig:stikov_substrate} and displayed them in Figure \ref{fig:single_fibres}. The top panel of the figure exhibits the cross-sections of the outer and inner surfaces of the fibre, along with the cross-sections of their diameters. The bottom panel shows the diameter distribution of each axon. We observed that, regardless of the tortuosity of the fibre trajectory, the diameter distribution of both the inner and outer diameters of all three cases was centred around the target diameter.

\section{Discussion}

Over the last 20 years, Monte-Carlo diffusion simulations have been used to optimise DW-MRI data acquisition protocols and validate microstructure models. Nevertheless, doubts have been raised regarding the accuracy of the simple geometries used to construct the diffusion substrates.

In this work, we introduced CACTUS, a novel framework to produce numerical substrates mimicking white matter tissue with rich microstructural features, closely matching the desired input parameters. Those parameters include the target distribution for inner and outer axon radii, g-ratio, orientation dispersion, fibre crossings, curvature, and packing density. The high versatility of CACTUS is founded on its efficient computational implementation and its mathematical formulation divided into three algorithmic steps (substrate initialisation, joint fibre optimisation, and fibre radial growth) composed of various competing terms controlling different substrate parameters.

To generate the substrates, we introduced a new algorithm to initialise fibre bundles with a target mean degree of orientation dispersion. Moreover, we introduced a novel capsule-based parametrisation for optimising fibre structures. Compared to circle parametrisations~\citep{Close2009, Ginsburger2019}, the capsule parameterisation requires fewer parameters, reducing the complexity of the optimisation problem. We adapted the cost functions inspired by~\citet{Close2009,Ginsburger2019} for capsules and provided analytical derivatives, making the optimisation faster and computationally more efficient. Finally, we proposed the fibre radial growth algorithm, which increases the fibre packing density in white matter substrates.

CACTUS was able to enhance the complexity of the fibre microstructure. In particular, our results showed CACTUS can produce substrate with fibre volume fraction beyond the 75\% previously achieved. CACTUS reached high fibre volume fractions, up to 95\% in its substrates (Table \ref{tab:dispersion_crossings}). Moreover, it consistently reached fibre volume fractions superior to 90\% at all the various levels of bundle dispersion and crossing angles (Table \ref{tab:dispersion_crossings} and Figure  \ref{fig:substrate_crossings}).

In the single bundle case, the fibre volume fraction was the highest at 94.7\% when fibres were aligned and decreased to 90.8\% with increasing mean angular dispersion. Conversely, the fibre volume fraction remained consistently around approximately 93\% in the two-bundle cases, regardless of the crossing angle. However, we note that the packing complexity of substrates with a single bundle and two bundles crossing differs. The former mimics the spatial arrangement of thousands of fibres with different crossing angles, which may produce more empty pockets between fibres and less densely packed substrates.

Another important feature of CACTUS is that it can create substrates with statistical characteristics informed by histological data. Indeed, we can closely adhere to the target statistics of axon volume fraction, myelin volume fraction, and g-ratio reported in histological studies~\citep{stikov2015vivo} (see 
 Figure~\ref{fig:stikov_substrate}). In all cases, the difference between the target and obtained substrate properties was lower than $2\%$ (see Table \ref{tab:dispersion_crossings}). Notably, CACTUS is the first tool incorporating the g-ratio as a target characteristic and successfully matching it for large-scale substrates.

Also, CACTUS has the capability to generate substrates with a targeted radii distribution. In our experiments, the approximation of the target distribution improves as substrate size increases, as illustrated in Figure \ref{fig:substrate_sizes}, underscoring the importance of generating large substrates. Furthermore, we have the availability to measure fibre geometry accurately. For instance, as seen in Figure \ref{fig:single_fibres}, the generated fibres have a non-constant longitudinal radius and non-circular cross-sections. Despite the tortuous trajectories of the fibres, the diameter distribution remains centred around the target mean outer (inner) diameter of $1.5 \mu m$ ($1.1\mu m $). Additionally, the diameter distribution presented replicates the diameter variations observed in 3D synchrotron images~\citep{andersson2020axon}, including longitudinal changes and a lack of skewness.

Finally, while previous works were able to achieve substrate sizes between $(30 \mu m)^3$ and $(100 \mu m)^3$, CACTUS demonstrated a substantial improvement in the generation of larger substrates~\citep{Callaghan2020, Ginsburger2019}. As shown in Figure \ref{fig:substrate_sizes}, CACTUS generated substrate sizes ranging from $(50\mu m)^3$ to $(500 \mu m)^3$, all with up to a $95\%$ fibre volume fraction. Our tool's ability to generate larger substrate sizes is advantageous for Monte-Carlo diffusion simulations in DW-MRI as it has been shown in previous studies~\citep{rafael2020robust}, that substrate sizes larger than $(200 \mu m)^3$ can reduce the sampling bias caused by smaller substrate sizes, potentially leading to more accurate DW-MRI numerical simulations
~\cite{rafael2020robust,romascano2018voxel}. In addition, the ability to generate large substrate sizes is advantageous as DW-MRI modelling is moving towards incorporating more microstructure features such as somas, astroglia, and vascularity~\citep{schneider2021structure,dyer2017quantifying,lin2018modelling}. This makes the generation of large substrates essential for capturing these additional features and moving towards more accurate and comprehensive microstructure imaging.

\subsection{Limitations and future work }

Although CACTUS incorporates complex microstructural features required to mimic some of the most relevant white matter geometrical properties, it still requires fibre-modelling assumptions to reduce the computational burden. Also, CACTUS generates substrates with characteristics resembling those from healthy white matter, but generating pathological tissue requires additional work, which we reserve for future studies. 

Additionally, CACTUS focuses solely on generating white matter fibre structures. However, its capacity to generate large substrate sizes expands the potential for including other tissue components in future studies, such as astrocytes, oligodendrocytes, microglia, and capillaries.

Finally, although CACTUS output substrates are suitable for simulators like the MCDC~\citep{rafael2020robust}, a thorough analysis is necessary to comprehend the influence of mesh quality, like the number of triangles, on the DW-MRI signals generated by Monte-Carlo simulation. Such analysis is crucial for developing computationally viable simulations.

\subsection{Applications beyond diffusion MR}

The applications of CACTUS are not limited to studying white matter microstructure using DW-MRI. For instance, it can be applied in DW-MRI studies outside the brain~\citep{adelnia2019diffusion}, where muscle fibres are organised into fascicles. The microscopic arrangement of muscle fibres can vary between different muscle groups, regions of the same muscle, and multiple pathological conditions \citep{berry2018relationships}.
Moreover, the fibre meshes generated by CACTUS could be used in other applications, like Polarized Light Imaging (PLI)~\citep{amunts2019dense,menzel2015jones}, a technique used to infer the local fibre orientation in histological brain sections based on the birefringent properties of the myelin sheaths. The limitations of the birefringence PLI model were investigated in~\citep{menzel2015jones} by generating synthetic PLI data from a hexagonal bundle of straight parallel cylindrical fibres. Although a more general fibre constructor was recently proposed for validating 3D-PLI techniques~\citep{amunts2019dense}, the white matter substrates generated in our study could provide more realistic geometries for conducting similar studies.

\section{Conclusion}

The generation of realistic substrates is critical for validating DW-MRI models, as it allows researchers to simulate and analyse the effect of microstructural changes on the DW-MRI signal. 

In this work, we introduced CACTUS, a novel framework for generating axonal-like substrates with predefined geometrical features of interest. Our experiments show that CACTUS can generate white matter substrates with the desired spatial dimensions, fibre radii, g-ratio, non-circular cross-sections, tortuous trajectories, smooth surfaces, predefined inter-fibre angles and fibre dispersion. Notably, the generated fibre substrates reached up to 95\% fibre volume fraction, the highest density reported in the literature to date, in agreement with previous histology studies. We also generated the large substrates/voxels of up to $(500 \mu m)^3$, with dimensions similar to or higher than those used in preclinical MRI scanners, reducing the gap between numerical and real voxel sizes.

In conclusion, the CACTUS substrate generator tool presented in this study has the potential to advance white matter microstructure modelling. It provides a versatile and customisable platform for generating fibre substrates with quantifiable geometrical characteristics. It is open-source and accessible to the broader research community at \url{http:/cactus.epfl.ch}, facilitating the validation and comparison of current and future DW-MRI models.

\section{Acknowledgments}

This work is supported by the Swiss National Science Foundation under grants $205320\_175974$ and $205320\_204097$. We acknowledge access to the facilities and expertise of the CIBM Center for Biomedical Imaging, a Swiss research centre of excellence founded and supported by Lausanne University Hospital (CHUV), University of Lausanne (UNIL), Ecole Polytechnique Federale de Lausanne (EPFL), University of Geneva (UNIGE) and Geneva University Hospitals (HUG). Erick J. Canales-Rodríguez was supported by the Swiss National Science Foundation (Ambizione grant $PZ00P2\_185814$).

We want to express our gratitude to Prof. Nikola Stikov and Dr. Jennifer Campbell for generously donating the histology images from their previous work~\citet{stikov2016vivo}. We thank Dr. Thomas Yu for his help proofreading this manuscript.

\section{Data availability}

The substrate meshes generated in this work and the source code are available at \url{http://cactus.epfl.ch}.

\section{Author contribution}

\bibliographystyle{Frontiers-Harvard} 

\bibliography{main}

\end{document}